\documentclass[preprintnumbers,prd,twocolumn,aps,showpacs,amsmath,amssymb,floats]{revtex4-1}
\usepackage{graphicx}
\usepackage{epsfig}
\usepackage{bm}
\usepackage{amsfonts}
\usepackage{amssymb}
\usepackage{subfloat,subfig}
\usepackage{times}
\usepackage{natbib}
\usepackage[titletoc,title]{appendix}
\usepackage{textcomp}
\usepackage{natbib}
\usepackage[normalem]{ulem}

\usepackage[dvips]{color}

\def\lapp{\mathrel{\rlap{\raise.5ex\hbox{$<$}}
                    {\lower.5ex\hbox{$\sim$}}}}
\def\gapp{\mathrel{\rlap{\raise.5ex\hbox{$>$}}
                    {\lower.5ex\hbox{$\sim$}}}}

\begin{document}
\title{Impact of Sommerfeld enhancement on helium reionization via WIMP dark matter}

\author{Bidisha Bandyopadhyay}
\email{bbandyopadhyay@astro-udec.cl}
\author{Dominik R.G. Schleicher}
\email{dschleicher@astro-udec.cl}
\affiliation{Departamento de Astronom\'ia, Facultad Ciencias F\'isicas y Matem\'aticas, Universidad de Concepci\'on, Av. Esteban Iturra s/n Barrio Universitario, Casilla 160-C, Concepci\'on, Chile}

\begin{abstract}
 Dark matter annihilation can have a strong impact on many astrophysical processes in the Universe. In the case of Sommerfeld-enhanced annihilation cross sections, the annihilation rates are enhanced at late times, thus enhancing the potential annihilation signatures. We here calculate the Sommerfeld-enhanced annihilation signatures during the epoch of helium reionization, the epoch where helium becomes fully ionized due to energetic photons. When considering the upper limits on the energy injection from the CMB, we find that the resulting abundance of He$^{++}$ becomes independent of the dark matter particle mass. The resulting enhancement compared to a standard scenario is thus 1-2 orders of magnitude higher. For realistic scenarios compatible with CMB constraints, there is no significant shift in the epoch of helium reionization, which is completed between redshifts $3$ and $4$. While it is thus difficult to disentangle dark matter annihilation from astrophysical contributions (active galactic nuclei), a potential detection of dark matter particles and its interactions using the Large Hadron Collider (LHC) would allow one to quantify the dark matter contribution.\end{abstract}

\maketitle
\section{Introduction}

In the standard model of cosmology, dark matter comprises one of the dominant components in the energy budget of the Universe, amounting to about $28\%$ according to Planck data \cite{Planck}. Our knowledge of structure formation requires dark matter to be cold, with the standard interpretation thus considering weakly interacting massive particles (WIMPs), with particles' masses in the range of $10-1000$~GeV. If the observed dark matter abundance is to be explained via thermal relics, the latter translates into a constraint on the annihilation cross section of $\langle \sigma_{\chi} v\rangle\sim3\times10^{-26}$~cm$^3$~s$^{-1}$ \cite{Bertone04}.\\ 

The resulting annihilation of dark matter can give rise to a number of astrophysical implications, which can potentially help to  discover or constrain the true nature of dark matter. Comparing predictions of annihilation signals in the Galactic Center with observed fluxes in particular in the gamma-ray regime have led to strong upper limits particularly for dark matter candidates in the MeV range, including lower bounds on the dark matter particle mass \cite{Ahn05a, Ahn05b}, the s-wave annihilation cross section \cite{Schleicher09a} and the annihilation rate into positrons \citep{Beacom05}. Similar upper bounds have been inferred from the microwave excess observed with WMAP \citep{Hooper}.\\ 

The Galactic Center is particularly well-suited for such dark matter probes, as the enhanced dark matter density will naturally boost the annihilation rate. In fact, while there is evidence for various types of emission coming from that region, it is still a matter of ongoing debate whether the latter could be due to astrophysical effects. This includes the $511$ keV line emission \cite{Jean, Weidenspointner, BoehmHo}, as well as the detection of GeV photons \cite{Boer2005}, microwave photons \cite{Hooper} and positrons \cite{Cirelli}. The Alpha Magnetic Spectrometer (AMS) on the International Space Station (ISS) recently confirmed with unprecedented precision \cite{Aguilar}, previous measurements from the PAMELA \cite{Picozza} and FERMI \cite{Gehrels} experiments showing a clear rise in the positron fraction above energies of $10$~GeV \cite{Adriani}, which have been tested against pulsar and dark matter models \cite{Cholis}. \\

At early cosmic times the mean dark matter density in the Universe is high and this leads to potentially observable effects. This is already relevant during the epoch of recombination, where dark matter annihilation may increase the available energy input and affect both the temperature and polarization power spectra of the  CMB \cite{Chen,Padmanabhan}. Specifically, it may cause an increased ionization fraction during recombination thereby causing an enhancement in the polarization signal at large angular scales \cite{Natarajan}. Similarly, dark matter particularly in the MeV mass range has been shown to potentially influence the epoch of reionization, where the Universe has been reionized by the first sources of light,  with possible contributions from self-annihilating dark matter. The latter has been shown by \cite{Mapelli06, Ripamonti07}, including the derivation of upper limits for the dark matter annihilation cross section \cite{Schleicher08}.\\ 

However, also the more massive WIMPs have been shown to produce potentially relevant astrophysical effects. For instance, in environments of high dark matter densities, including the Galactic Center or the epoch of reionization, dark matter annihilation can act as a power source inside the stars and therefore alter their stellar evolution as well as the lifetimes \citep[see e.g.][]{Iocco08, Spolyar08, Scot09, Smith12}. In the context of such models, it is conceivable that some of the first stars that formed in the Universe are still existing at the present day. It should however be noted that constraints and upper limits can be derived based on the observed gamma-ray background and the duration of reionization \citep{Schleicher09b}.\\

 As recently shown by \cite{Bandyopadhyay}, the same may occur during the epoch of helium reionization, where highly energetic particles produced in dark matter annihilation events contribute to the transition from the singly-ionized state to the doubly-ionized state of helium. As shown in that study, the main contribution to this effect arises around redshifts of $2-4$, where the clumpiness in the dark matter distribution is strongly increasing, thus considerably enhancing the annihilation signal. The epoch of helium reionization has been probed through the absorption lines of quasars, including Q0302003 \cite{Jakobsen,Heap}, HS 1700+6416 \cite{Davidsen,Fechner}, HE 23474342 \cite{Reimers1997,Kriss,Smette,Shull}; and HST observations of PKS 1935-692 \cite{Anderson}, SDSS J2346-0016 \cite{ZhengCh,ZhengKr,ZhengMe}, Q1157+3143 \cite{Reimers2005} and SDSS J1711+6052 \cite{ZhengMe}. These data suggest helium ionization to occur between redshifts of $2.2$ and $3.8$.\\
 
 While this study concerning the effect on helium reionization considered generic WIMP models with the cross section based on dark matter relics, the effective cross section at a given redshift may however be time-dependent, depending on the annihilation mechanism. While annihilation through a p-wave channel with the cross-section scaling as $v^2$ \cite{Diamanti} leads to a suppression in the s-wave contribution, processes like the Sommerfeld Enhancement, with the cross sections scaling as $v^{-\alpha}$ for $\alpha>0$, can potentially enhance annihilation effects at late times. The latter could make the effect of dark matter annihilation more relevant at late cosmological epochs, including the epoch of helium reionization. We adopt here the Sommerfeld enhancement model by \cite{Feng}, which gives rise to an increased effective annihilation cross section at late times, while maintaining the same initial cross section, and therefore fulfilling the constraint based on the relic dark matter density. \\
 
 We present the overall framework of our investigation in section~\ref{2}, providing a summary of the previous formalism outlined by \cite{Bandyopadhyay}. The specific effects due to dark matter annihilation, including our treatment of clumping and Sommerfeld enhancement, are then described in section~\ref{3}. In this section, we also provide a generalized definition of the dark matter clumping factor, which accounts for potential local variations of the Sommerfeld enhancement. The results are presented in section~\ref{4} and discussed in section~\ref{5}. \\

\section{The Ionization Equation}\label{2}
In this investigation, we build up on the framework previously developed by \cite{Bandyopadhyay}, which will be extended here to include the effect of Sommerfeld enhancement. In the following, we summarize the main equations with a brief discussion of the main physical processes. Our modified treatment of the dark matter will then be presented in the next section.\\

As a minimal model for early cosmic times, we  consider here the ionization effects caused by collisional ionization of He$^{+}$, energy injection by dark matter annihilation and the recombination to He$^{+}$. In addition at late times, we  also consider the effect of energetic photons from high redshift quasars. As has been shown by \cite{Furlanetto2008} they are the major contributors leading to the complete ionization of Helium in the IGM. The rate at which the mean ionized fraction of Helium evolves is given by
\begin{eqnarray}
 \frac{d\bar{x}_{\rm He^{++}}}{dt}&=&k_{QSO}+k_{DM}+(1.-\bar{x}_{He^{++}}) n_e \beta_{He^{+}}\nonumber \\
 &-&\bar{C}\alpha_A (T)n_e\bar{x}_{He^{++}}. \label{eq1}
\end{eqnarray}
The second term describes the contribution from the annihilation of dark matter and is studied in detail in the next section. The first term here considers the ionization due to photons from quasars. If $\dot{N}(L_B)=2\times10^{55}$~s$^{-1}L_B/(10^{12}\,L_\odot)$ is the total number of photons with energy greater than the double-ionization threshold of helium ($54.4$~eV) emitted per second by a quasar with luminosity $L_B$ and $\frac{d\phi}{dL_B}$ is the quasar luminosity function (QLF), then we have \cite{Furlanetto2008}
\begin{equation}
k_{QSO}=\int dL_B \frac{\dot{N}(L_B)}{\bar{n_{He}}}\frac{d\phi}{dL_{B}}.  \label{eq2}
\end{equation}
The third and fourth term in eqn.~\ref{eq1} are the collisional and radiative recombination terms, respectively. The fraction of ionized helium is expressed as $\bar{x}_{\rm He^{++}}=n_{He^{++}}/n_{He}$ while the number density of electrons is $n_e$. The collisional ionization coefficient $\beta_{He^+}$ and the recombination coefficient $\alpha_A$ \cite{Cen} depend on the matter temperature and evolve as
\begin{eqnarray}
 \beta_{He^+} &=& 5.68 \times 10^{-12}T^{1/2}e^{-631515/T} \nonumber \\
 &\times&\left[1+\left(\frac{T}{10^5}\right)^{1/2}\right]^{-1} \mathrm{cm}^3\ \mathrm{s}^{-1}, \label{iocoeff}\\
 \alpha_A(T)&=&3.36 \times 10^{-10}T^{-1/2}\left(\frac{T}{10^3}\right)^{-0.2}\nonumber \\
 &\times&\left[1+\left(\frac{T}{10^6}\right)^{0.7}\right]^{-1} \mathrm{cm}^3\ \mathrm{s}^{-1}, \label{recocoeff}
\end{eqnarray}
while for the baryonic clumping factor ($\bar{C}=\langle \rho^2 \rangle / \langle \rho\rangle^2$) we have chosen the following adaptation at various redshifts $z$:
\begin{equation}
 \bar{C}=
 \begin{cases}
  1 &\text{($z>15$)}\\
  1+\frac{15-z}{9} &\text{($6<z<15$)}\\
  3 &\text{($z<6$)} .
 \end{cases}
\end{equation}
We therefore adopt a uniform baryon density for  $z > 15$ {\it i.e.} before a significant amount of structure formation has taken place. A value of $\bar{C}=3$ is adopted for $z < 6$ which has also been considerd by \cite{Furlanetto2008}. A linear interpolation is then pursued in between. One can note that the baryonic clumping factor used here is smaller in comparison to those used in studies for hydrogen reionization \cite{Ciardi, Schleicher08}. This is a valid assumption since helium reionization is driven by rare objects like quasars and therefore proceeds on larger spatial scales. The processes described here are important at different cosmological epochs and have a characteristic influence on the evolution of ionized helium at different times. Since the collisional ionization and recombination coefficients are strongly dependent on the IGM temperature, we have used the RECFAST code \cite{Seager} to follow the gas temperature for $z>15$. We then further assume that the IGM temperature rises to $\rm T =10000$ K during hydrogen reionization.

\begin{figure*}[htb!]
\subfloat[]{ \includegraphics[height=2.2in,width=2.5in,angle=-90]{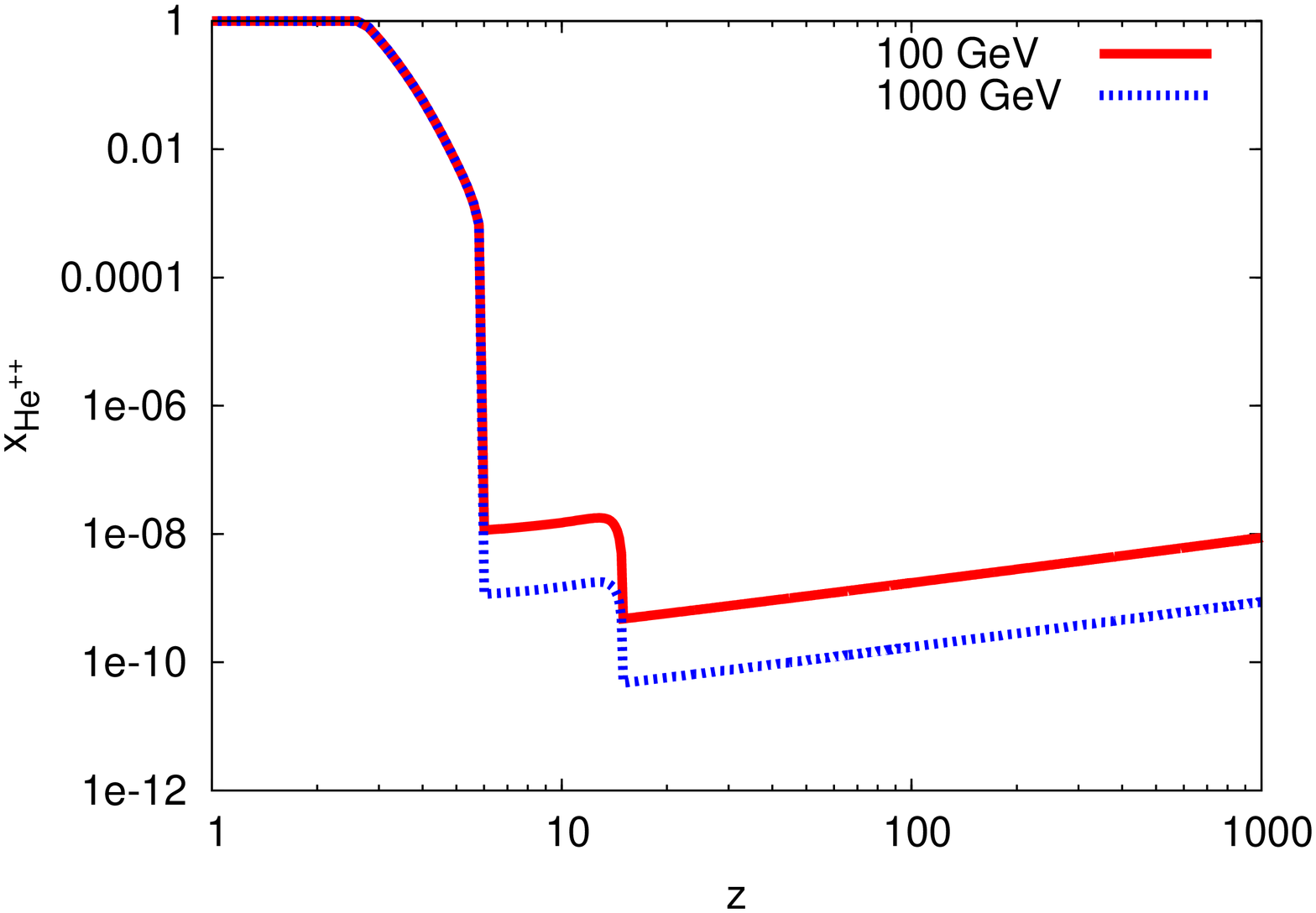} \label{gra1}}
\subfloat[]{ \includegraphics[height=2.2in,width=2.5in,angle=-90]{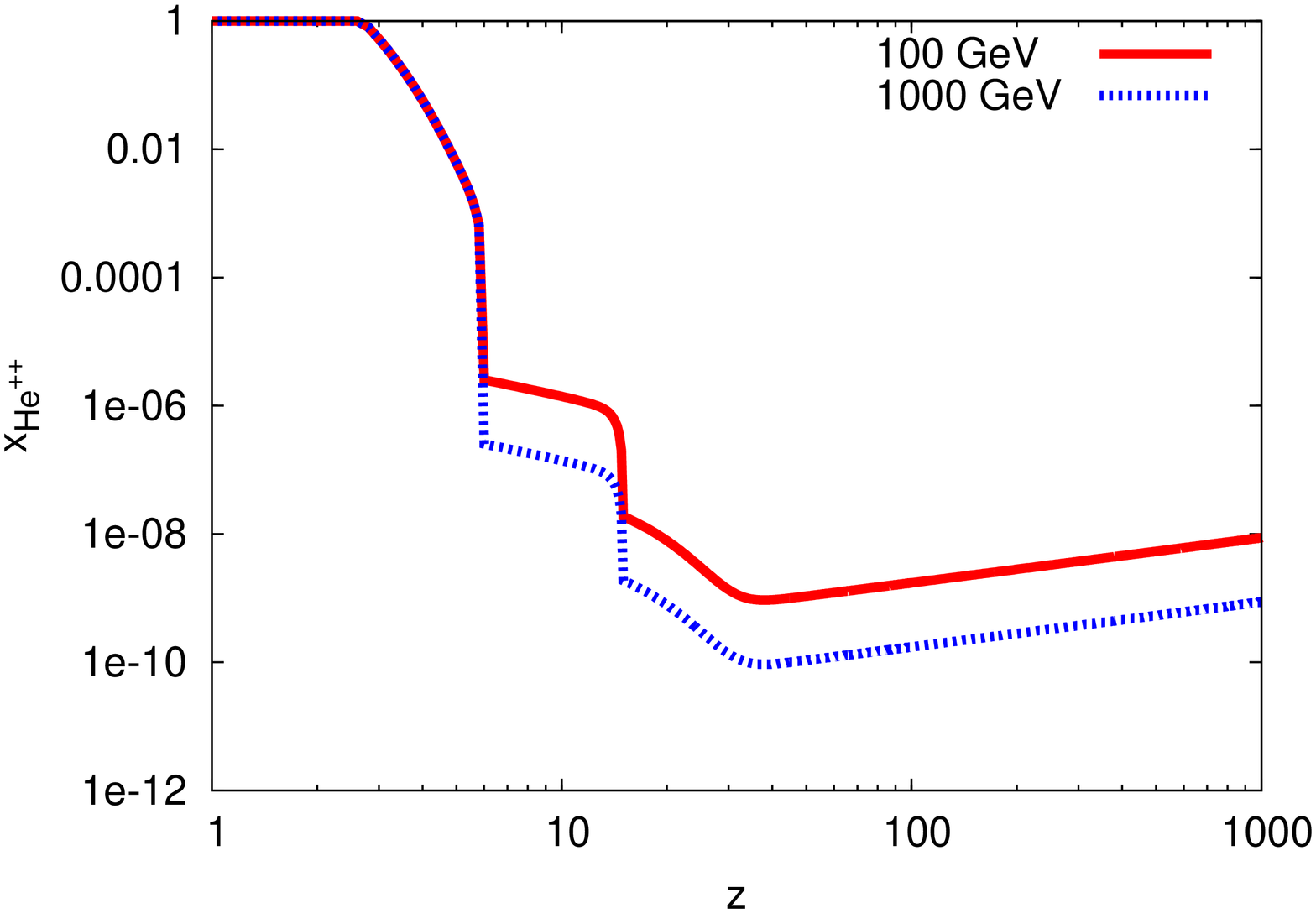}  \label{gra2}}
\subfloat[]{ \includegraphics[height=2.2in,width=2.5in,angle=-90]{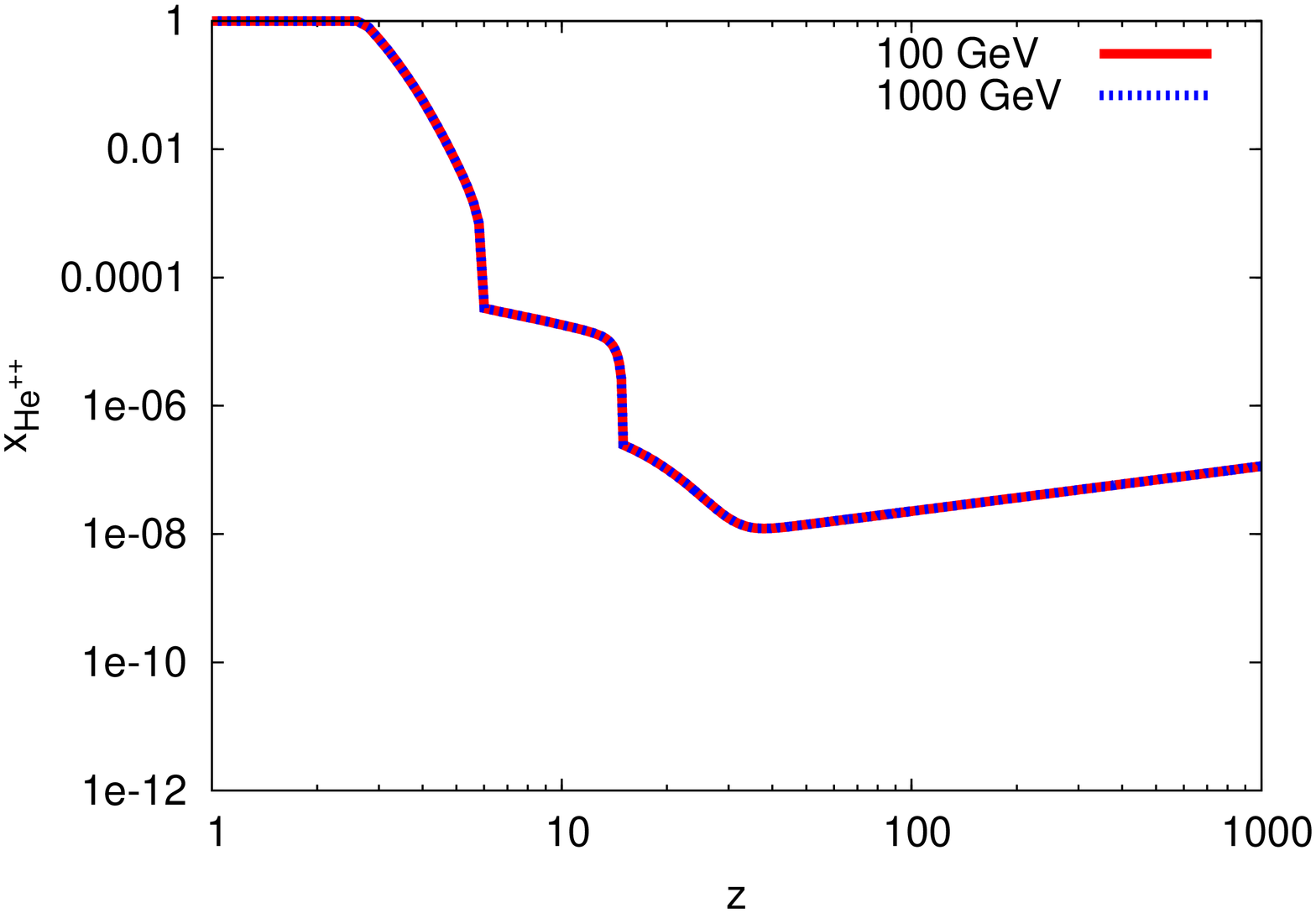} \label{gra3}}
 \begin{center}
\caption{Evolution of He$^{++}$ in the presence of quasars and dark matter annihilation for $100$ GeV and $1000$ GeV cold dark matter in different scenarios: [\ref{gra1}] when the dark matter is uniformly distributed, [\ref{gra2}] when there is clumping due to structure formation and [\ref{gra3}] when in addition there is Sommerfeld Enhancement.} \label{cap1}
\end{center}
\end{figure*}

\section{Effect of Dark Matter Annihilation}\label{3}
To investigate the effect of dark matter annihilation on helium reionization including Sommerfeld enhancement, we adopt here a typical framework which is employed to study the impact of dark matter annihilation on hydrogen reionization \cite{Schleicher08}. This approach going back to \cite{Furlanetto06} assumes that the dark matter particles annihilate into high-energy photons, which allows to determine the energy fractions going into heating, ionization and excitation. While \cite{Furlanetto06} employed the simplified approach by \cite{Shull85}, we generalized the latter approach to include also the double-ionization of helium \cite{Bandyopadhyay} adopting the formalism of \cite{Dalgarno}. Direct annihilation of dark matter particles to photons with thermal cross-sections however violate the HESS results \cite{Cline}. We thus assume annihilation to intermediate electron-positron pairs which annihilate to photons. We adopt the on-the-spot approximation of energy injection, as used by \cite{Oldengott}, into the IGM effectively assuming that the injection timescale is short compared to the timescale of cosmic evolution.\\

 We note that of course more detailed energy injection mechanisms have been worked out in the past \cite{Slatyer13, Valdes10, Valdes, Evoli, Slatyer16a, Slatyer16b}, which can potentially lead to additional energy losses. For instance, \cite{Evoli} have calculated the energy going into different processes for specific dark matter models, including a 10-GeV bino-like neutralino, a heavy dark matter candidate of rest mass 1~TeV that pair annihilates into muons, and a 200~GeV wino-like neutralino that pair annihilates into W$^+$W$^-$ pairs. We also refer to very recent work by \cite{Slatyer16a} which updated a lot of the microphysical processes. Others like \cite{Valdes10} have worked out the energy cascade processes leading to energy depositions in the IGM.\\

 Considering the overall uncertainties regarding the nature of dark matte and the various injection processes, we here intend however to not investigate specific models, but we rather focus on a generic dark matter scenario just with and without the Sommerfeld enhancement. As described further in section~\ref{3b}, the Sommerfeld enhancement is treated here in a generic way, without considering possible resonances, that would diminish the Sommerfeld effect and bring the results closer to the standard scenario that was already investigated. Such a specific study can be pursued when there is a strong motivation for one particular dark matter candidate, but is not the focus of the work employed here. We note that an accurate treatment of the ionization processes of helium is ensured through the formalism of \cite{Dalgarno}, which allows us to precisely determine the amount of energy going into helium double-ionization.\\

In our approach, we thus write the ionization term due to dark matter annihilation $k_{DM}$ as
\begin{eqnarray}
k_{DM}&=&\eta_2\chi_i\langle \sigma v \rangle\left(\frac{m_p c^2}{E_{\rm He^{+}, ion}}\right)\nonumber\\
&\times&\frac{\Omega_{\chi}\rho_{\rm crit}}{m_{\chi}} \left(\frac{\Omega_{\chi}}{\Omega_b}\right)(1+z)^3 \label{eq3}
\end{eqnarray}
where $\langle\sigma v \rangle$  is the thermally averaged cross section for annihilation and is a model dependent parameter. $\chi_i$ is the fraction of energy that goes into ionizing He$^{+}$ and is defined as $\chi_i=E/E_{ph}$, where $E$ is the energy that goes into ionizing the atom while $E_{ph}$ is the energy of the incoming particle. {The details of the energy cascade effect is in a way embeded in the parameter $\chi_i$ as mentioned in \cite{Dalgarno}}. The total energy required for ionization is $E=N \times E_{\rm ion}$ where $N$ is the total number of ionizations and $E_{\rm ion}$ is the ionization threshold energy. From \cite{Dalgarno}, the mean energy per ion pair $W=E_{ph}/N$ is parametrically given as $W=W_0(1+Cx^{\alpha})$, with parameters $W_0$, $C$ and $\alpha$ that depend on the gas composition. On substitution, one obtains $\chi_i=E_{ion}/W$. For photons with energies greater than $1$~keV and a cosmological mixture of  hydrogen and helium gases the value of the constant parameters are $W_0=16400$~eV, $C=11.7$ and $\alpha=1.05$. For He$^+$ ionization it is assumed that $\eta_2 = 8/0.24$, while the ionization threshold is $E_{\rm He^{++}, ion}=54.4$~eV.
\begin{figure*}[htb!]
\subfloat[]{ \includegraphics[height=3.0in,width=2.5in,angle=-90]{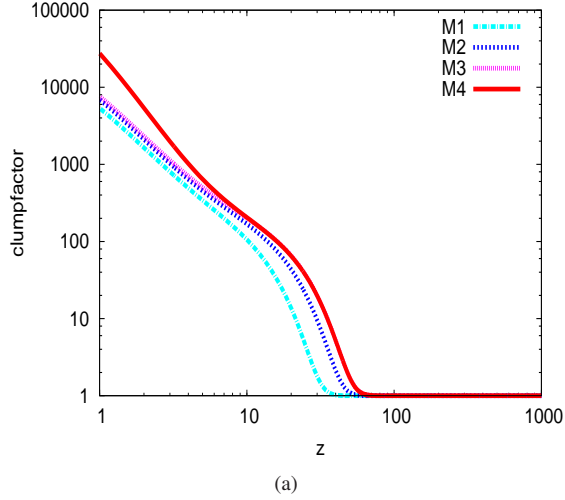} \label{gra6}}
 \begin{center}
\caption{Evolution of the clumping factor for different mass ranges ($M_1 \rightarrow 10^6-10^{12}$~M$_\odot$, $M_2 \rightarrow 10^4-10^{12}$~M$_\odot$, $M_3 \rightarrow 10^{-12}-10^6$~M$_\odot$ and $M_4 \rightarrow 10^{-12}-10^6$~M$_\odot$ with $M_{cut}=10^6 M_{\odot}$).} \label{cap2}
\end{center}
\end{figure*}

Here we consider three models (a) uniform density of dark matter with the cross section being the same as at freeze out,  (b) clumping in dark matter due to the formation of large scale structures and (c) the cross section with a Sommerfeld enhancement factor. Since the exact dark matter particle masses are not known, we have chosen a $100$ GeV and a $1000$ GeV mass candidate for this exploration. We adopt here the bound obtained from Planck on the annihilation cross section {\it i.e.} $\langle\sigma v \rangle_0\sim3 \times 10^{-26}$~cm$^3$~s$^{-1}$ in eqn.~\ref{eq3} \cite{Steigman}.

\begin{figure*}[htb!]
\subfloat[]{ \includegraphics[height=2.2in,width=2.5in,angle=-90]{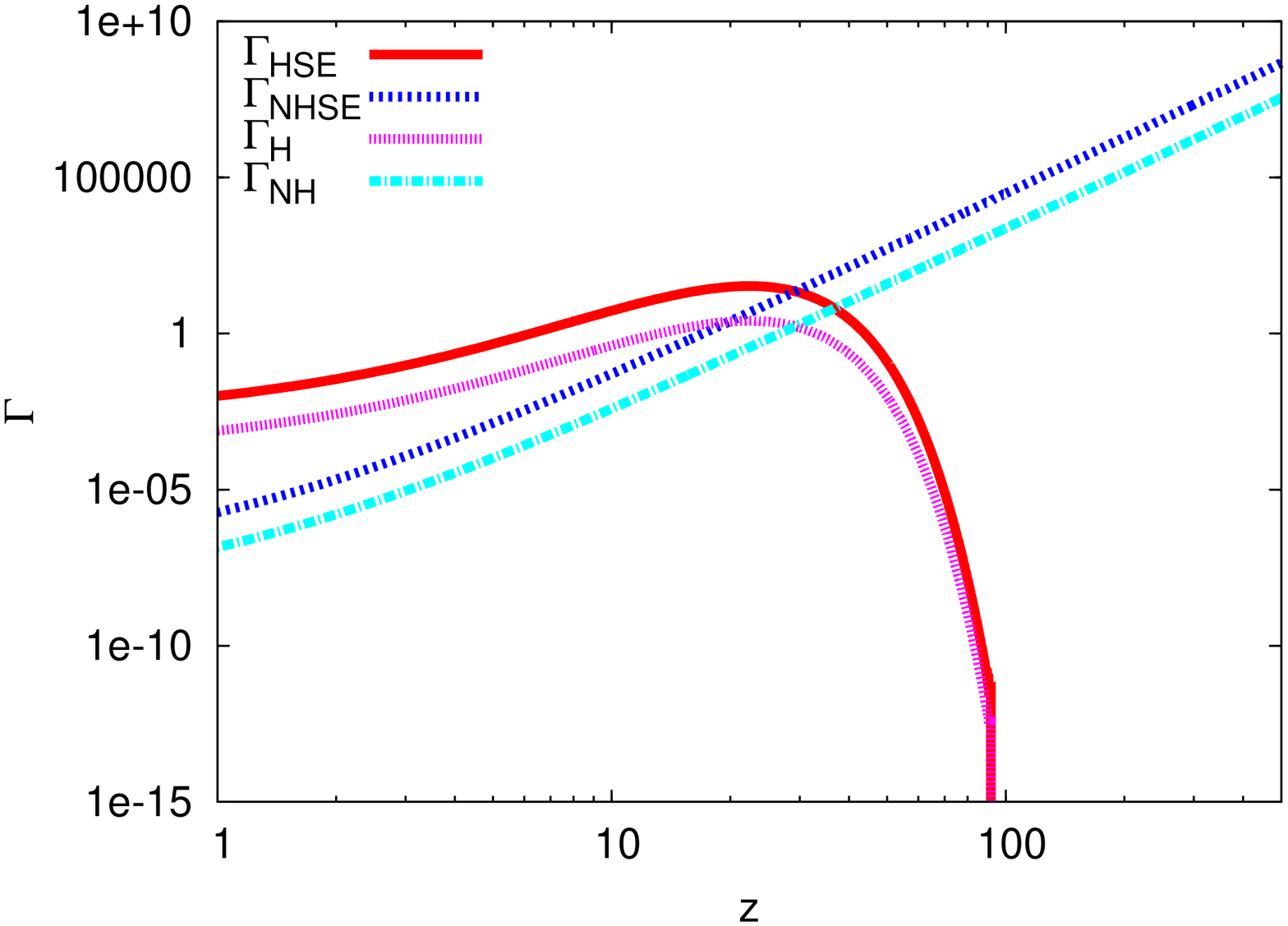} \label{gammacomp}}
\subfloat[]{ \includegraphics[height=2.2in,width=2.5in,angle=-90]{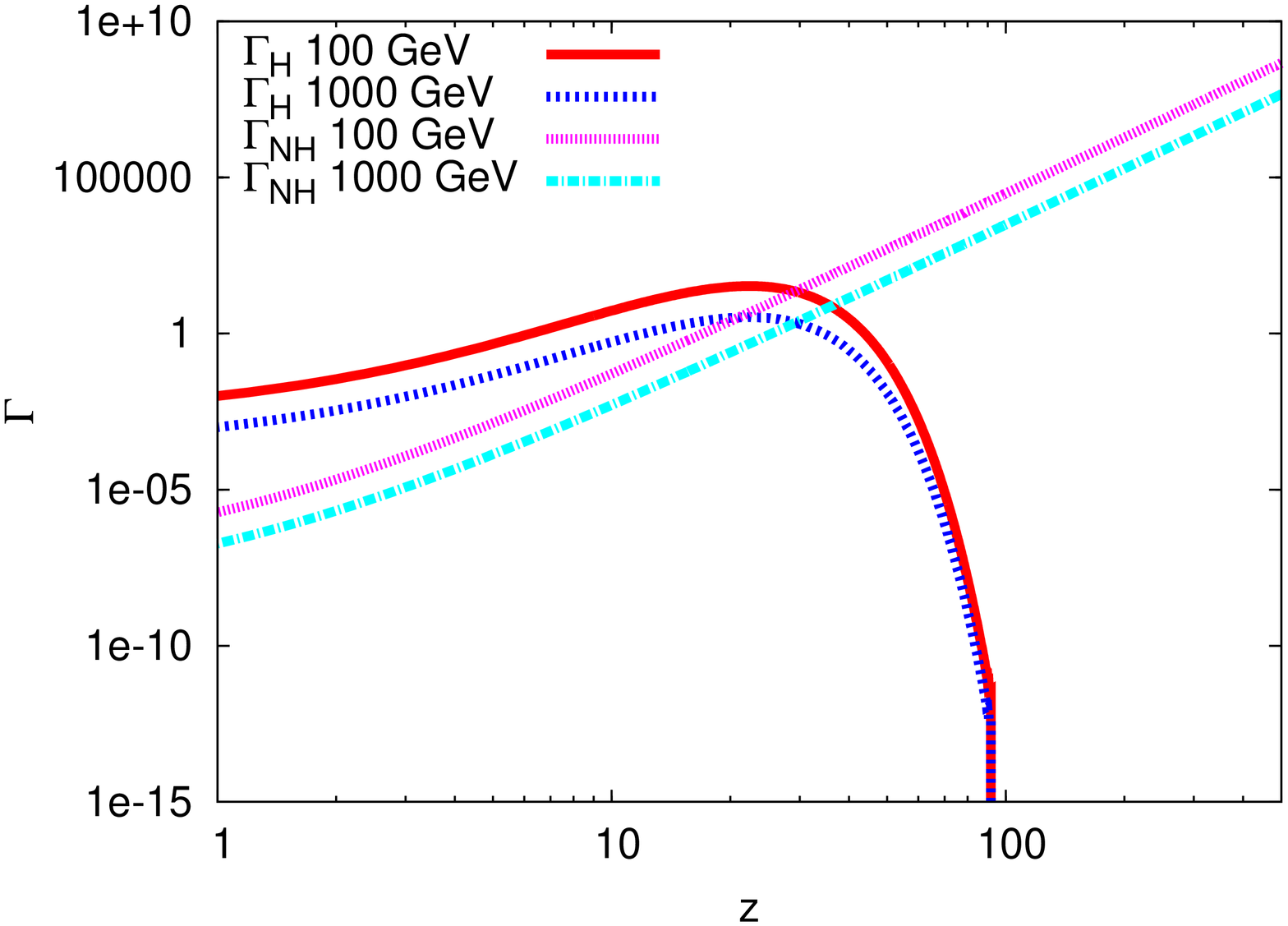}  \label{gamma1001000}}
\subfloat[]{ \includegraphics[height=2.2in,width=2.5in,angle=-90]{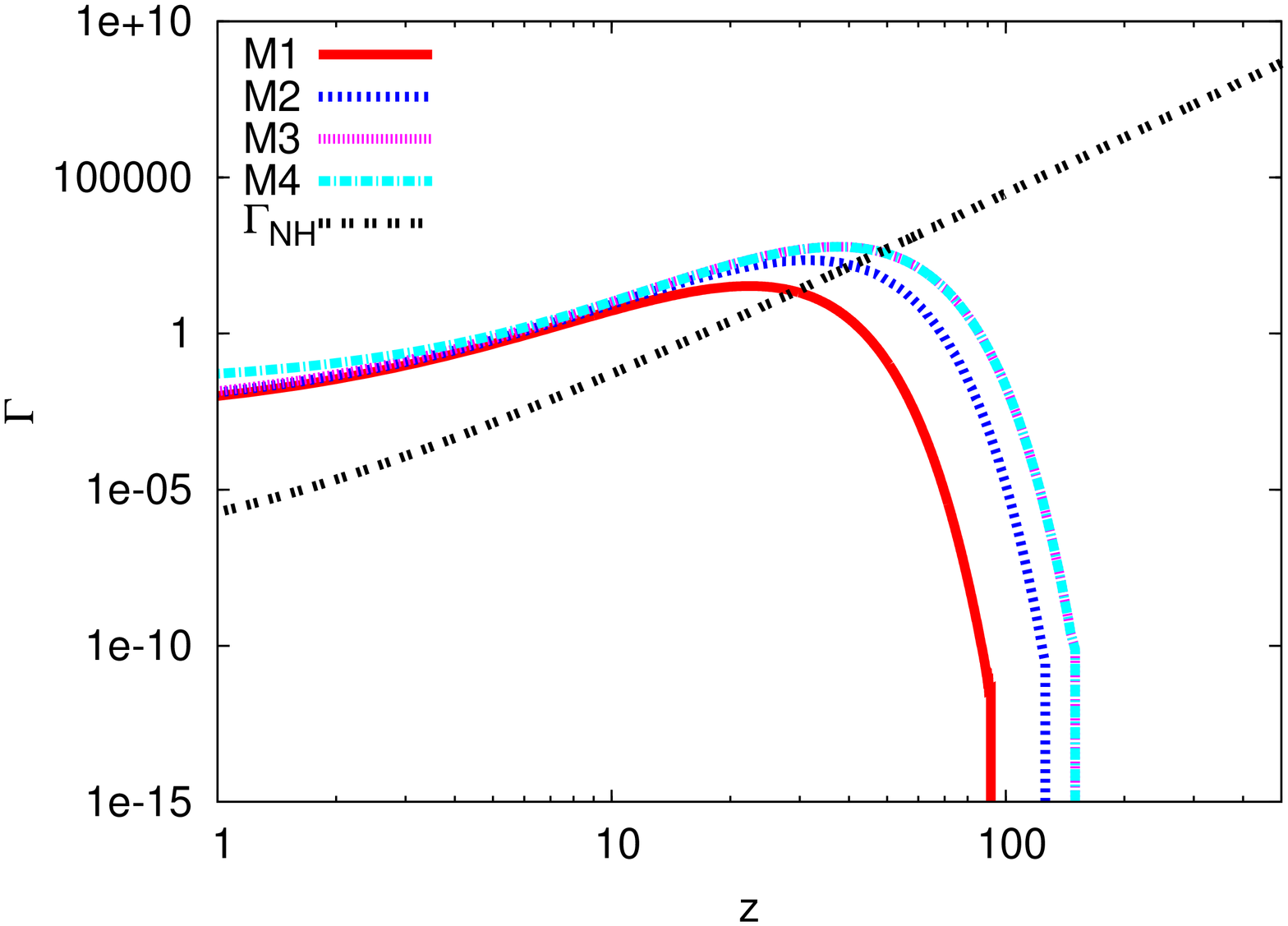} \label{gammaSE100}}
 \begin{center}
\caption{Evolution of the annihilation rates within the halo $\Gamma_{H}$ and the background $\Gamma_{NH}$ for [\ref{gammacomp}] the case of annihilation with ($\Gamma_{HSE}$ and $\Gamma_{NHSE}$) and without ($\Gamma_{H}$ and $\Gamma_{NH}$) Sommerfeld annihilation for $100$ GeV dark matter candidate, [\ref{gamma1001000}] the case of annihilation with Sommerfeld enhancement for $100$ GeV and $1000$ GeV dark matter particles and the annihilation with Sommerfeld enhancement for different halo mass ranges ($M_1 \rightarrow 10^6-10^{12}$~M$_\odot$, $M_2 \rightarrow 10^4-10^{12}$~M$_\odot$, $M_3 \rightarrow 10^{-12}-10^6$~M$_\odot$ and $M_4 \rightarrow 10^{-12}-10^6$~M$_\odot$ with $M_{cut}=10^6 M_{\odot}$).} \label{gamma}
\end{center}
\end{figure*}

\subsection{Clumpy dark matter}\label{3a}
The dark matter annihilation rate depends on the square of its density. At early cosmic times we consider a uniform density of dark matter, but later dark matter starts clumping on account of large scale structure formation. This non-uniform distribution in density is accounted for by a clumping factor which describes the enhancement in the annihilation rate within the high density regions of dark matter halos. This clumping factor depends both on the number of such high density regions and the density profile within such halos. The cross section in eqn.~\ref{eq3} can be modified to include the clumping factor and can be written as $\langle\sigma v \rangle = C_{halo}\langle\sigma v \rangle_0$ where $C_{\rm halo}=1$ implies a uniform density.\\

Unlike our previous study \cite{Bandyopadhyay} where we had considered NFW, Moore and Burkert profiles for the dark matter density distribution, we focus here on the NFW profile for illustrative purposes \cite{NFW1996,NFW1997,Salucci}, as it is also considered to be more realistic. We would however obtain qualitatively similar results when considering a different profile. We adopt here a similar approach to calculate the clumping factor as done by Cumberbatch et al. \cite{Cumberbatch}. The clumping factor can be described as the ratio of the annihilation rate within the halos ($\Gamma_{halo}$) to the annihilation rate within the smooth background ($\Gamma_{smooth}$), i.e.
\begin{equation}
 C_{halo}(z)=1+\frac{\Gamma_{halos}(z)}{\Gamma_{smooth}(z)}, \label{enclum}
\end{equation}
where
\begin{eqnarray}
&\Gamma_{smooth}(z)=&\frac{1}{2}\frac{\langle \sigma_{\chi}v \rangle}{m_{\chi}^2}\bar{\rho}^2_{\chi}(z),\nonumber\\
 &\Gamma_{halos}(z)=&(1+z)^3 \nonumber \\
 &\times&\int_{M_{min}}^{M_{max}}dM\frac{dn}{dM}(M,z)R(M,z).
\end{eqnarray}
The annihilation rates here imply the number of dark matter particles annihilating per unit time. Here, $\bar{\rho}_{\chi}$ is the mean dark matter density, $\frac{dn}{dM}$ is the mass function which determines the number of halos in a unit mass range (we have used the Press-Schechter mass function \cite{Press}) and $R(M,z)$ is the rate of annihilation within a halo of mass $M$ and redshift $z$, which is given by
\begin{equation}
 R(M,z) = \frac{1}{2}\frac{\langle \sigma_{\chi}v \rangle}{m_{\chi}^2}\int_{r=0}^{r_{vir(M,z)}}\rho^2(r)4\pi r^2dr.
\end{equation}
For a halo with an NFW like density profile, the density distribution is a continuous function of the radius $r$ and is given by
\begin{equation}
 \rho(r)=\frac{\rho_s}{(r/r_s)^{\gamma}[1+(r/r_s)^{\alpha}]^{(\beta-\gamma)/\alpha}}, \label{eq5}
\end{equation}
with $\alpha=\gamma=1$ and $\beta=3$.

\begin{figure*}[htb!]
\subfloat[]{ \includegraphics[height=3.0in,width=2.5in,angle=-90]{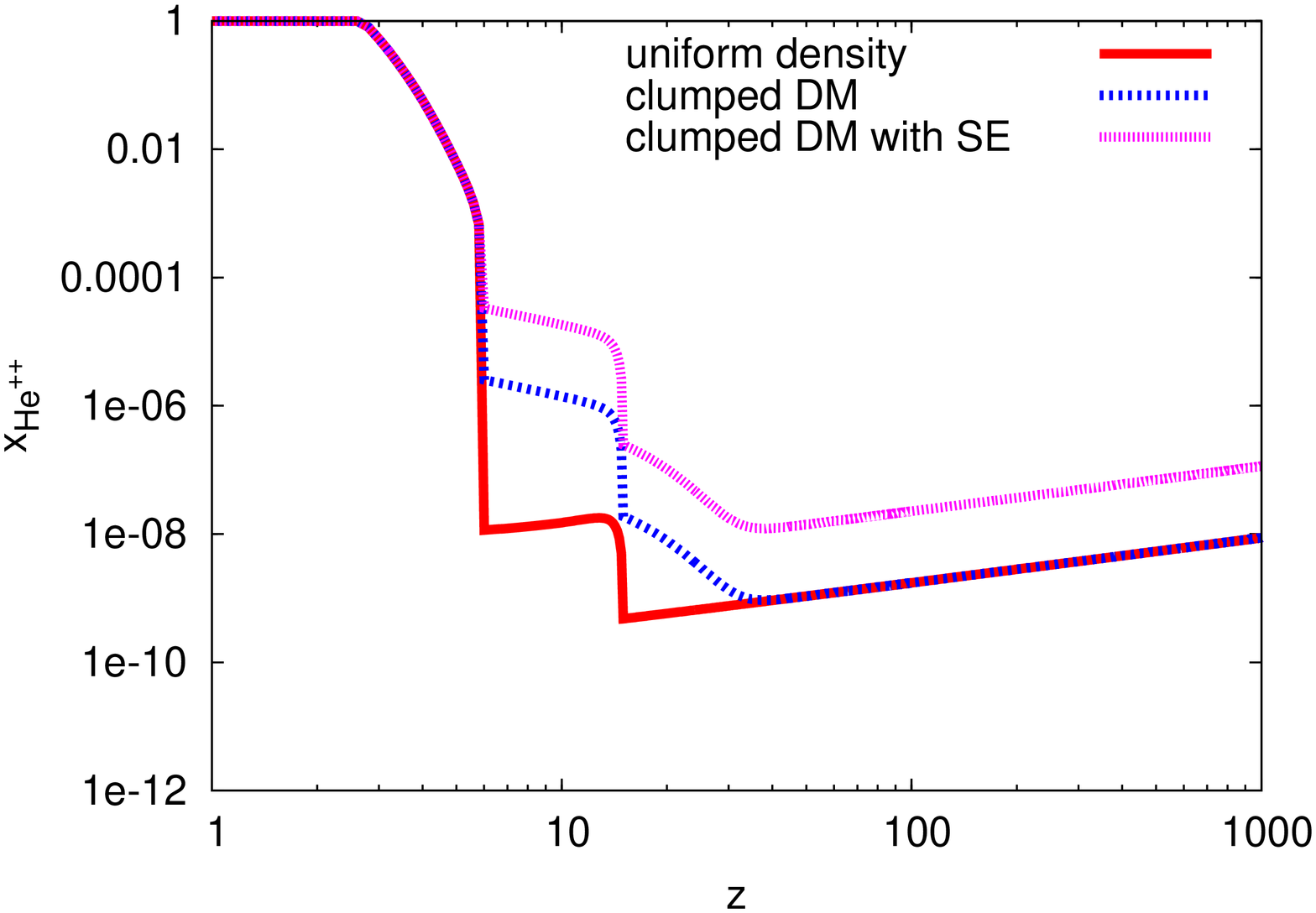} \label{gra7}}
\subfloat[]{ \includegraphics[height=3.0in,width=2.5in,angle=-90]{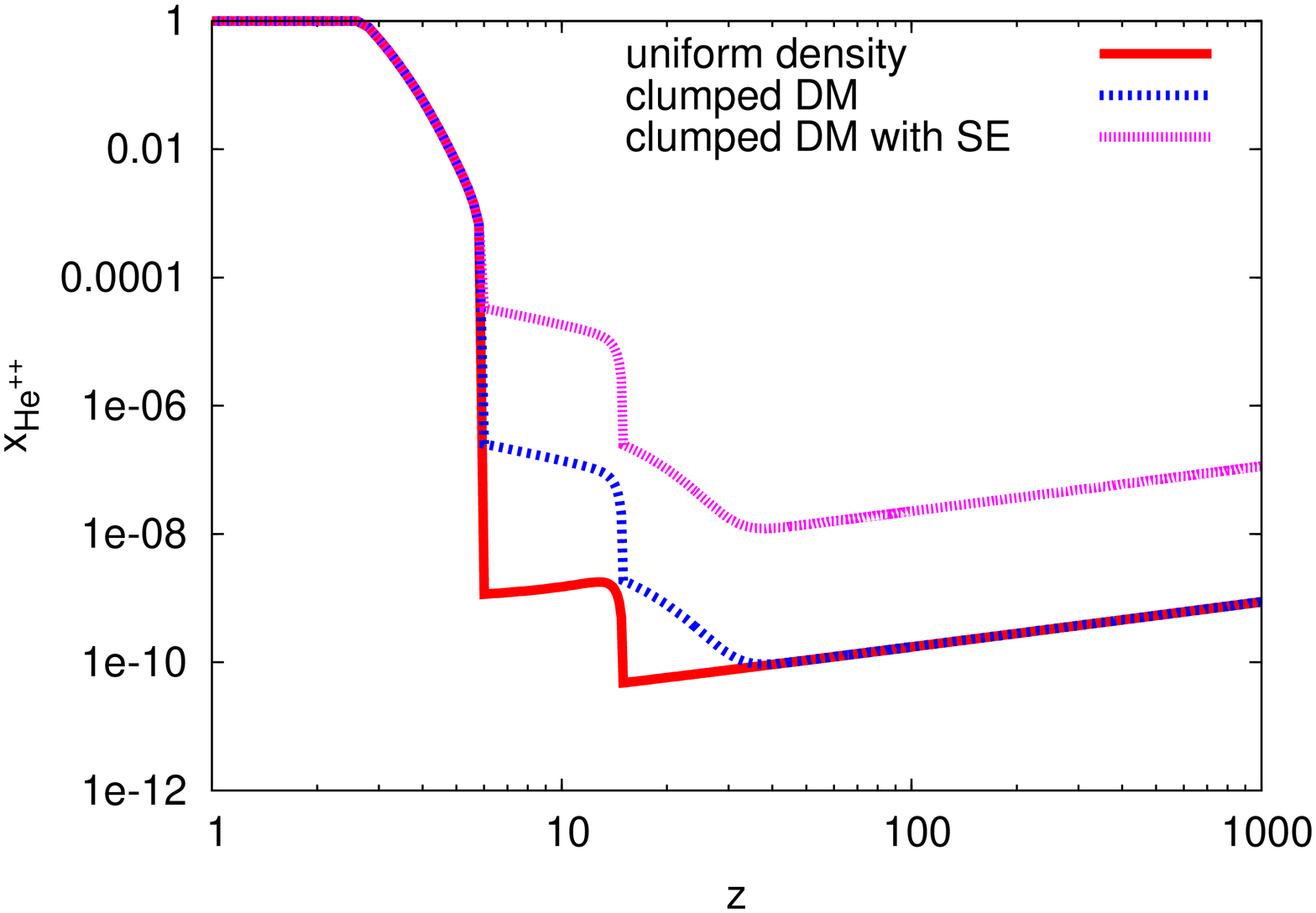}  \label{gra8}}
 \begin{center}
\caption{Evolution of He$^{++}$ in the presence of quasars and a dark matter annihilation uniform background, clumpy dark matter and with Sommerfeld Enhancement for [\ref{gra7}] $100$ GeV and [\ref{gra8}] $1000$ GeV dark matter candidates.} \label{cap3}
\end{center}
\end{figure*}

\begin{figure*}[htb!]
 \subfloat[]{ \includegraphics[height=3.0in,width=2.5in,angle=-90]{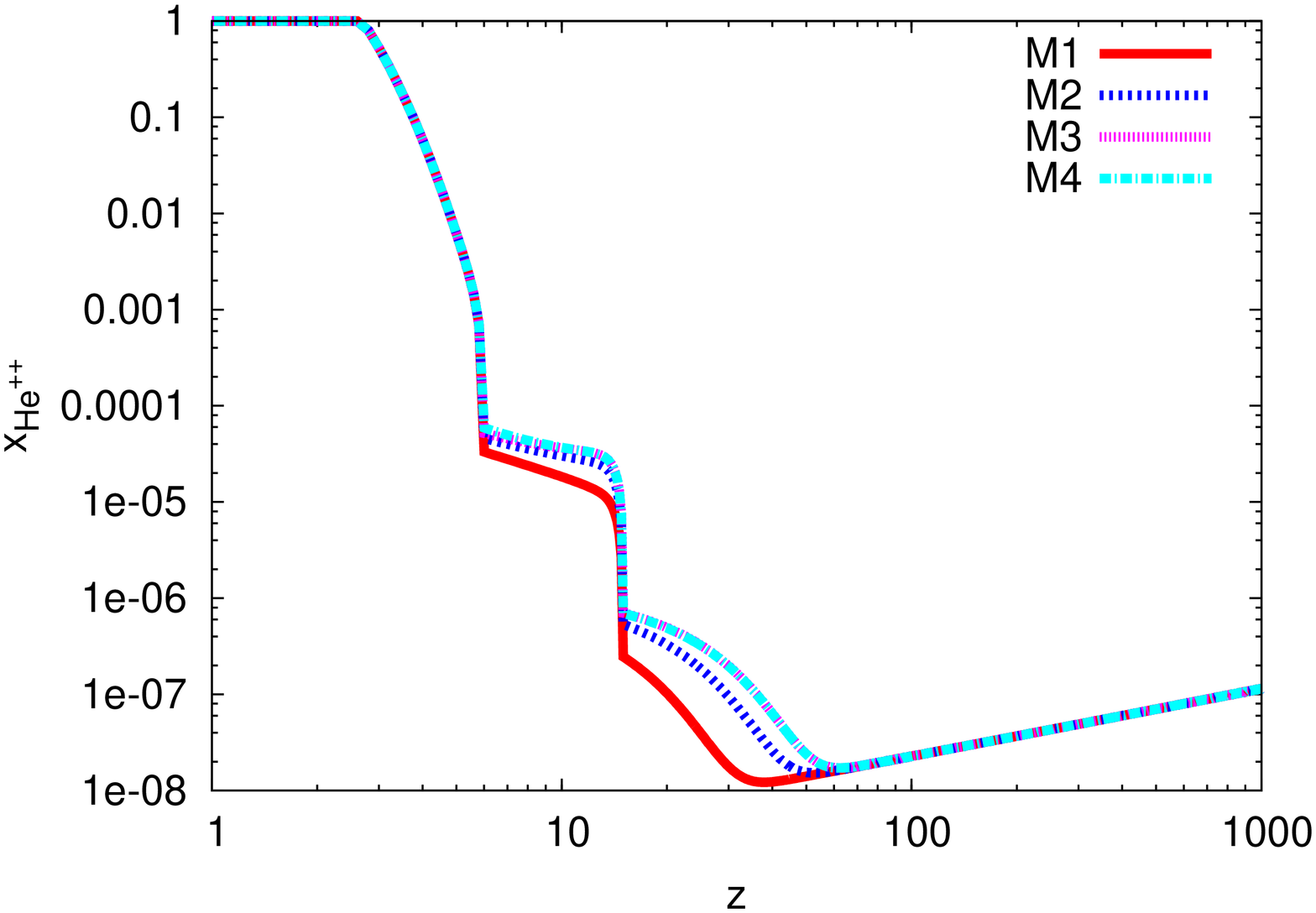} \label{gra9}}
 \subfloat[]{ \includegraphics[height=3.0in,width=2.5in,angle=-90]{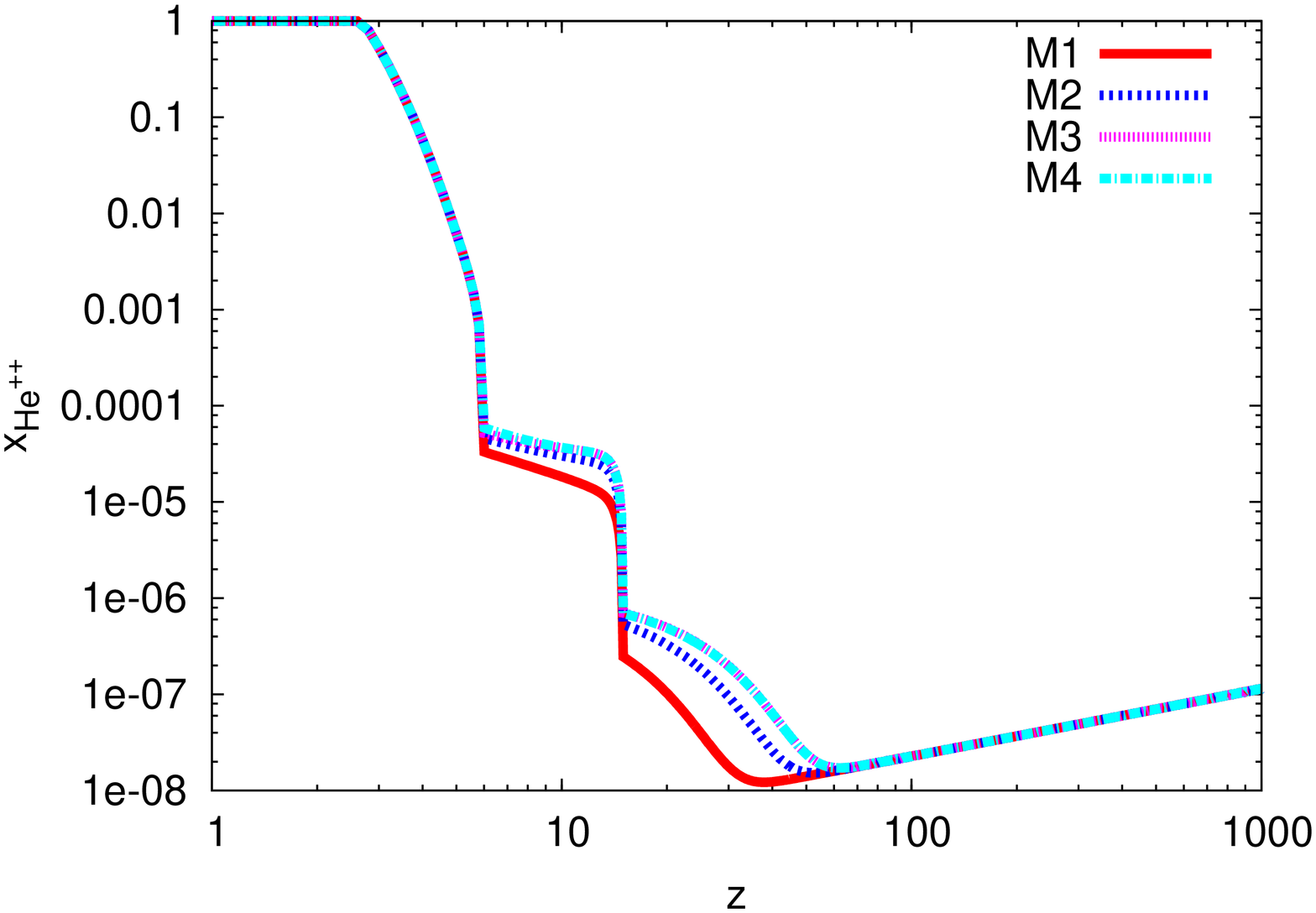}  \label{gra10}}
  \begin{center}
 \caption{Evolution of He$^{++}$ in the presence of quasars and dark matter annihilation for [\ref{gra8}] $100$ GeV and [\ref{gra10}] $1000$ GeV dark matter particles for various mass ranges ($M_1 \rightarrow 10^6-10^{12}$~M$_\odot$, $M_2 \rightarrow 10^4-10^{12}$~M$_\odot$, $M_3 \rightarrow 10^{-12}-10^6$~M$_\odot$ and $M_4 \rightarrow 10^{-12}-10^6$~M$_\odot$ with $M_{cut}=10^6 M_{\odot}$).} \label{cap4}
 \end{center}
 \end{figure*}

\subsection{Sommerfeld enhanced cross section}\label{3b}
Now to introduce a more realistic particle physics model, one needs to generalize the expression for the dark matter annihilation cross section which is dependent on the relative velocity of the dark matter particles. The cross-section can be modified to include the Sommerfeld Factor $S$ such that $\langle \sigma_{\chi}v \rangle=C_{SH}\bar{S}( \sigma_{\chi}v )_0$ \cite{Feng}.  If one considers the annihilation of dark matter particles $\chi$ into some intermediate particles $\phi$ which subsequently decay to standard model particles, then one can consider the tree level cross section dependence as 
\begin{equation}
 (\sigma_{\chi} v)_0 \sim \frac{\pi\alpha_{\chi}^2}{m_{\chi}^2},
\end{equation}
where for a typical weak scale $m_{\chi}$ the thermal relic density implies a weak coupling $\alpha_{\chi}\approx 0.05~[m_{\chi}/2~\rm TeV]$. In the case of $\alpha_{\chi} \gg v$ \cite{Feng}, we have the original Sommerfeld factor 
\begin{equation}
 S=\frac{\pi \alpha}{v}.
\end{equation}
The mean Sommerfeld Factor $\bar{S}$ for a non-relativistic velocity distribution is then
\begin{equation}
\bar{S}(M)\simeq \frac{x_{0}^{3/2}}{2\sqrt{\pi}N}\int_0^{v_{max}}Sv^2e^{-x_0 v^2/4}dv,
\end{equation}
where $x_0=2/v_0^2$ and $N=\rm erf(z_s/\sqrt{2})-(2/\pi)^{1/2}z_se^{-z_s^2/2}$ with $\rm z_s\equiv v_{max}/v_0$. In these expressions $v_{0}$ is the average velocity. The maximum velocity of particles within the halos can be approximated with the escape velocity of particles as $v_{\rm max H}=\sqrt{\frac{2G M}{R}}$ while $v_{0H}=\sqrt{\frac{G M}{R}}$ is the circular velocity within the halo. The mean velocity $v_{0NH}$ of particles in the smooth background can be approximated to be the average thermal velocity given by $v_{0NH}\approx\sqrt{\frac{kT_{\chi}}{m_{\chi}}}$, where $T_{\chi}$ is the dark matter temperature. The dark matter temperature is a redshift dependent quantity and can be expressed in terms of the kinetic decoupling temperature ($T_{kD\chi}$) and the redshift of decoupling ($z_{D\chi}$) of dark matter as $T_{\chi}=T_{kD{\chi}}\left(\frac{1+z}{1+z_D\chi}\right)$The maximum velocity $v_{max NH}$ in principle corresponds to the speed of light, but considering the typical velocity distribution, we make a cut-off at ten times the thermal velocity. In the general case, one thus expects that the enhancement within the halo $\bar{S}_H$ may be different from the one in the smooth background $\bar{S}_{NH}$.\\ 

In the following, these velocities will be evaluated based on the halo mass $M$, assuming that at a given redshift, halos have a characteristic size following from their mass and virial density. With these assumptions, the clumping factor in eqn.~\ref{enclum} is modified to take into account the Sommerfeld enhancement via
\begin{equation}
 C_{SH}(z)=1+\frac{\bar{S}_{\rm H, av}\Gamma_{halos}(z)}{\bar{S}_{\rm NH, av}\Gamma_{smooth}(z)}, \label{modclump}
\end{equation}
where the $\bar{S}_{\rm H, av}$ is defined via\begin{equation}
\bar{S}_{\rm H, av}=\frac{\int_{M_{min}}^{M_{max}}dM\frac{dn}{dM}(M,z)R(M,z)\bar{S}(M)}{\int_{M_{min}}^{M_{max}}dM\frac{dn}{dM}(M,z)R(M,z)}.
\end{equation}
The Sommerfeld enhancement factor cancels out in obtaining the clumping factor in case the contribution from the halo and the uniform component are equal, which in particular can happen in the saturated regime. CMB observations provide a stringent bound on the amount of energy injected at the epoch of decoupling, which then provides an upper bound on the maximal enhancement \cite{Galli, Slatyer} given as
\begin{equation}
 S_{max}<\frac{120}{f}\left(\frac{m_{\chi}}{\rm TeV}\right).
\end{equation}
This leads to the saturation of this effect for small values of the velocity. This bound is much lower than the one given by the particle physics models. It should be noted that the way the clumping factor has been defined here ensures that there is no double counting of the annihilation rates. In the absence of structures or halo formation, the clumping factor $C_{SH}=1$. Taking into account the modifications for the Sommerfeld Enhancement, eqn.~\ref{eq3} is  rewritten as
\begin{eqnarray}
k_{DM}&=&\eta_2\chi_i \bar{S}_{\rm NH, av}C_{SH}(\sigma_{\chi} v)_0\left(\frac{m_p c^2}{E_{\rm He^{+}, ion}}\right)\nonumber\\
&\times&\frac{\Omega_{\chi}\rho_{crit}}{m_{\chi}} \left(\frac{\Omega_{\chi}}{\Omega_b}\right)(1+z)^3. 
\end{eqnarray}
We have not considered here the possible presence of resonances in the Sommerfeld enhancement, as such a case is highly model-dependent and essentially leads to a suppression in the s-wave annihilation cross section. This rather diminishes the contribution of the Sommerfeld enhancement at late epochs. Such potential cases may thus be expected to lie inbetween the calculations considered here and the calculations without Sommerfeld enhancement that we reported earlier \cite{Bandyopadhyay}. 

\section{Results}\label{4}
The evolution of the ionized fraction of helium as a function of redshift is shown for various cases in Fig.~1. In Fig.~\ref{gra1} and Fig.~\ref{gra2} one can see this evolution for a uniform dark matter density and for clumpy dark matter, respectively. In both  cases we have assumed a constant cross section ($\langle\sigma v \rangle\sim3 \times 10^{-26} cm^3s^{-1}$) for the $100$ GeV and $1000$ GeV dark matter particle masses. The difference in the ionized fraction for these masses clearly shows that the annihilation rate for a constant annihilation cross section is $\propto m_{\chi}^{-2}$. For the case of uniform dark matter density we observe that the ionized fraction initially falls with redshift which shows the decrease in dark matter number density due to the expansion of the Universe. Then, there is a sudden increase at a redshift of $15$ due to the increase in ionization temperature as a result of hydrogen ionization. Finally there is an increase in helium ionization due to the photons from quasars from about a redshift of $6$ which completely ionizes helium at a redshift of about $2.5$. On the other hand for clumpy dark matter we observe that there is an increase in helium ionization due to the clumping of dark matter due to the fact that the annihilation rate increases due to the increase in dark matter inhomogeneity during structure formation. We have assumed here a simple Press-Schechter mass function which shows an early peak in the annihilation rate compared to other models like Sheth-Tormen or the one by Watson et.el. \cite{Watson} as seen in the works of \cite{Diamanti}. This however will not affect our results much due to Sommerfeld enhancement as we have shown that this effect reaches its saturation value even before the epoch of decoupling and thus will not lead to a significant change in Helium ionization fraction due to Sommerfeld enhancement by choosing a different mass function.\\

In Fig.~\ref{gra3} we observe that in case of Sommerfeld enhancement the increase in the ionized fraction is the same for both dark matter particle masses. This shows that the enhanced cross section and the energy injected per annihilation for the higher mass compensates for the decrease in the number density. Thus we realize that in case of Sommerfeld enhancement the factor $\bar{S}(\sigma_{\chi}v)(m_{\chi}c^2)/m_{\chi}^2$ remains constant and is independent of the particle mass. This is expected when the Sommerfeld effect has reached its saturation value in the uniform background as well as in the halos.\\

We also show the effect of dark matter particle masses and Sommerfeld enhancement on the clumping factor. Fig.~\ref{cap2} shows that the clumping factor is independent of the dark matter particle masses because the clumping factor for the 100 GeV and 1000 GeV particles overlap and thus are not shown separately in the plot. The clumping factor is also independent of the Sommerfeld enhancement implying that this factor is equal for both the halos ($S_H$) and the uniform background ($S_{NH}$) are equal, as the Sommerfeld Effect is saturated both in the halos as well as the uniform background. This happens because the CMB bound of Sommerfeld factor is more stringent than the one obtained from simplified particle physics models. In Fig.~\ref{cap2}, we also observe that the clumping factor depends on the range of the adopted halo masses and the concentration parameters.\\

In Fig.~\ref{gamma} we have plotted the annihilation rates within the halos and for the smooth background for different scenarios as a function of redshift. Fig.~\ref{gammacomp} shows this evolution for the cases of annihilation with and without Sommerfeld enhancement for $100$ GeV dark matter particle. It can be clearly seen that the ratio of the annihilation rates within the halo to the rate within the background in both the cases will be same. This confirms our result for the behavior of the clumping factor, which is defined as the ratio of these two quantities. This kind of a behaviour is possible only if the Sommerfeld effect reaches its saturation value both in the uniform and clumped regions much before the epoch of decoupling. The behaviour for the $100$ GeV and $1000$ GeV dark matter particles is very similar as shown in Fig.~\ref{gamma1001000}. While the annihilation rate for the $1000$ GeV dark matter candidate is reduced compared to the case of $100$ GeV candidate, the evolution of the ionized fraction of helium is the same in both cases, as the energy released per annihilation by a $1000$ GeV candidate is greater compared to a $100$ GeV candidate, thus compensating for the lower annihilation rate. In Fig.~\ref{gammaSE100} we show the annihilation rate within the halos for various mass ranges of the halos for the $100$~GeV dark matter particles. For comparison we have also plotted in the same figure the rate of annihilation for the background.\\

We also compare the evolution of the ionized fraction in Fig.~\ref{cap3} for the three scenarios adopting dark matter particle masses of $100$ GeV and $1000$ GeV. It can be clearly seen that for the scenario of clumping without Sommerfeld enhancement the ionized fraction is suppressed at all redshifts and the suppression is larger for the $1000$ GeV dark matter candidate. In Fig.~\ref{cap4} we have shown the evolution of the ionized fraction with Sommerfeld enhancement with various halo mass ranges and we observe that both dark matter particle masses lead to identical results for the various halo mass ranges.

\section{Discussion and Conclusion}\label{5}
We have investigated the effect of annihilation of clumpy dark matter with Sommerfeld enhanced cross sections on the epoch of helium reionization. Here we have used the Sommerfeld enhancement model of particle physics \cite{Feng} which suggests that the cross section of annihilation gets enhanced by a factor of $v^{-1}$, considering the upper bounds available from the CMB \cite{Galli, Slatyer}. Our results show a clear enhancement of the annihilation signal compared to the case without the Sommerfeld effect. In addition, we find that the resulting abundance of He$^{++}$ becomes independent of the mass of the dark matter particle, as the enhanced cross sections for the larger dark matter particle masses compensate for the lower number density of the particles. The independence of the dark matter particle mass results from adopting the limits on energy injection from the CMB. If instead we were to calculate the ionization based on particle physics models (which depends on the weak coupling constant), a mass dependence would likely be retained. The latter thus leads to a qualitative change on how dark matter annihilation affects helium reionization, and we expect that similar results would apply for the epoch of hydrogen reionization as well \citep[see e.g.][]{Mapelli06, Ripamonti07, Schleicher08}. Various annihilation channels can lead to the production of electron positron pairs but as shown by \cite{Feng} the $4-\mu$ channels is the most probable channel. Such uncertainities need to be taken into account for a more careful modeling.\\

We also find that the Sommerfeld enhancement saturates quite early, thus leading to the same relative enhancement of the annihilation rate both in the homogeneous background as well as in collapsed halos. Due to this early saturation, the clumping factor of the self-annihilating dark matter turns out to be the same as the case without Sommerfeld Enhancement. As in our previous study \cite{Bandyopadhyay}, the contribution of dark matter annihilation to helium reionization becomes most relevant between redshifts of $2-5$, where the contribution from quasars also rises significantly. Also in the presence of a Sommerfeld enhancement, it is thus still difficult to disentangle both contributions, requiring a rather precise census of energetic photons produced via AGN. Alternatively, if potential dark matter particles and their interactions are detected at the Large Hadron Collider (LHC), one may work out the resulting contributions to the helium reionization. As we find here, these would be independent of the dark matter particle mass, at least in the presence of Sommerfeld enhancement.\\ 

The calculations presented here of course assume that dark matter consists of WIMPs. While this is still considered as the main hypothesis to explain the abundance of dark matter, we note that the recent LIGO detections of black hole mergers \cite{Abbott16a, Abbott16b} have stimulated the debate whether primordial black holes may partly contribute to the dark matter density in the Universe \cite{Bird16}. Similarly, the larger diversity of dwarf galaxy rotation curves \cite{Oman15} and the ambiguous explanation of small objects like Hercules \cite{Blana} and Segue~1 \cite{Dominguez} remains a challenge within the current paradigm. We therefore expect that a clear detection of the dark matter particle, for instance via the LHC or as an astrophysical signature via the Cherenkov Telescope Array (CTA) will be necessary to uniquely identify the nature of dark matter.

\begin{acknowledgements}
BB thanks the German Academic Exchange Service (DAAD) for the funding of a collaborative visit via the program ''A New Passage to India'', and Prof. T.R. Seshadri for the support of this project. BB and DRGS thank for funding via ALMA-Conicyt (project code 31160001). DRGS further acknowledges funding via Fondecyt regular (project code 1161247), the "Concurso Proyectos Internacionales de Investigaci\'on, Convocatoria 2015" (project code PII20150171), Quimal (project code QUIMAL170001), the Anillo (project code ACT172033) and the BASAL Centro
de Excelencia en Astrofísica yTecnologías Afines (CATA) grant (project code PFB-06/2007).
\end{acknowledgements}


\begin{thebibliography}{5}
\bibitem{Adriani} Adriani, O. et al., 2009, Nature, 458, 607
\bibitem{Aguilar} Aguilar, M. et al., 2013, Phys. Rev. Lett. 110, 141102
\bibitem{Ahn05a} Ahn, K., Komatsu, E., 2005, PRD, 72 1301
\bibitem{Ahn05b} Ahn, K., Komatsu, E., 2005, PRD, 71, 1303
\bibitem{Anderson} Anderson S. F., Hogan C. J., Williams B. F., Carswell R. F., 1999, AJ, 117, 56
\bibitem{Abbott16a} Abbott, B.P. et al., 2016, PRL, 116, 061102
\bibitem{Abbott16b} Abbott, B.P. et al., 2016, PRL, 116, 241103
\bibitem{Bandyopadhyay} Bandyopadhyay B., Schleicher D. R. G., 2015, Phys. Rev. D, 92, 02
\bibitem{Beacom05} Beacom, J.F., Bell, N.F., Bertone, G., 2005, PRL, 94, 1301
\bibitem{Bertone04} Bertone, G., Hooper, D., Silk, J., 2004, Physics Reports 405, 279
\bibitem{Blana} Bla\~{n}a, M., Fellhauer, M., Smith, R., Candlish, G.~N., Cohen, R., Farias, J.~P., 2015, MNRAS, 446, 144
\bibitem{Bird16} Bird, S., Cholis, I., Mu\~{n}oz, J.B., Ali-Haimoud, Y., Kamionkowski, M., Kovetz, E.D., Raccanelli, A., Riess, A.G., 2016, PRL, 116, 201301
\bibitem{BoehmHo} Boehm C., Hooper D., Silk J., Casse M., and Paul J., Physical Review Letters 92, 101301 (2004), arXiv:astro-ph/0309686
\bibitem{Boer2005} Boer W. de, Sander C., Zhukov V., Gladyshev A.V., and Kazakov D.I., 2005, A\&A 444, 51, arXiv:astro-ph/0508617 
\bibitem{Cen} Cen R., 1992, ApJS, 78, 341C
\bibitem{Chen} Chen, X.-L., Kamionkowski, M., 2004, Phys. Rev. D, 70, 043502, arXiv:astro-ph/0310473
\bibitem{Cholis} Cholis, I.,  Hooper, D., 2013, Phys. Rev., D88, 023013, arXiv:1304.1840
\bibitem{Ciardi} Ciardi, B.; Ferrara, A.; White, S. D. M., 2003, MNRAS, 344, 7
\bibitem{Cirelli} Cirelli M., Franceschini R., and Strumia A., 2008, Nuclear Physics B 800, 204, 0802.3378
\bibitem{Cline} Cline J. M., Scott P., 2013, JCAP, 03, 044 
\bibitem{Cumberbatch} Cumberbatch, D.T., Lattanzi M. and Silk, J., 2010, PRD, 82, 10
\bibitem{Dalgarno} Dalgarno, A., Yan, M., Liu, W., 1999, ApJS, 125, 37
\bibitem{Davidsen} Davidsen A. F., Kriss G. A., Zheng W., 1996, Nature, 380, 47
\bibitem{Diamanti} Diamanti, R., Lopez-Honorez, L., Mena, O., Palomares-Ruiz, S., Vincent, A. C., 2014, JCAP, 1402, 017, arXiv:1308.2578
\bibitem{Dominguez} {Dom{\'{\i}}nguez}, R., {Fellhauer}, M., {Bla{\~n}a}, M., {Farias}, J.P., {Dabringhausen}, J., {Candlish}, G.N., {Smith}, R., {Choque}, N., 2016, MNRAS, 461, 3630
\bibitem{Fechner} Fechner C., Reimers D., Kriss G. A., Baade R., Blair W. P., Giroux M. L., Green R. F., Moos H. W., Morton D. C., Scott J. E., Shull J. M., Simcoe R., Songaila A., Zheng W., 2006, A\&Ap, 455, 91
\bibitem{Evoli12} Evoli, C.; Vald\'es, M.; Ferrara, A.; Yoshida, N., 2012, MNRAS, 422, 420
\bibitem{Evoli} Evoli, C.; Mesinger, A.; Ferrara, A., 2014, JCAP, 11, 024
\bibitem{Feng} Feng, Jonathan L., Kaplinghat, Manoj, Yu, Hai-Bo, 2010, Phys. Rev D, 82, 08
\bibitem{Furlanetto06} Furlanetto, Steven R.; Oh, S. Peng; Briggs, Frank H., 2006, Physics Reports, 433, 181
\bibitem{Furlanetto2008} Furlanetto, S.R. and Oh, S.P. \, 2008 \, ApJ \, 681 \, 1
\bibitem{Galli} Galli S., Iocco F., Bertone G., Melchiorri A., 2009, Phys. Rev. D, 80, 02 
\bibitem{Gehrels} Gehrels, N., Michelson, P., 1999, Astropart. Phys. 11, 
\bibitem{Heap} Heap S. R., Williger G. M., Smette A., Hubeny I., Sahu M. S., Jenkins E. B., Tripp T. M., Winkler J. N., 2000, ApJ, 534, 69
\bibitem{Hooper} Hooper D., Finkbeiner D.P., and Dobler G., 2007, Phys. Rev. D 76, 083012, 0705.3655 
\bibitem{Iocco08} Iocco, F., 2008, ApJ, 677, 1
\bibitem{Jakobsen} Jakobsen P., Boksenberg A., Deharveng J. M., Greenfield P., Jedrzejewski R., Paresce F., 1994, Nature, 370, 35
\bibitem{Jean} Jean P., Kn\"ödlseder J., Gillard W., Guessoum N., Ferri\'ere K., Marcowith A., Lonjou V., and Roques J.P.,2006, A\&A 445, 579, arXiv:astro-ph/0509298 
\bibitem{Kriss} Kriss G. A., Shull J. M., Oegerle W., Zheng W., Davidsen A. F., Songaila A., Tumlinson J., Cowie L. L., Deharveng J.-M., Friedman S. D., Giroux M. L., Green R. F., Hutchings J. B., Jenkins E. B., Kruk J. W., Moos H. W., Morton D. C., Sembach K. R., Tripp T. M., 2001, Science, 293, 1112
\bibitem{Mapelli06} Mapelli, M., Ferrara, A., Pierpaoli, E., 2006, MNRAS, 369, 1719
\bibitem{Natarajan} Natarajan, A., Schwarz, D. J., 2009, Phys. Rev. D. 80, 043529
\bibitem{NFW1996} Navarro J. F., Frenk C. S., White S. D. M., 1996, ApJ., 462, 563
\bibitem{NFW1997} Navarro J. F., Frenk C. S., White S. D. M., 1997, ApJ., 490, 493
\bibitem{Oldengott} Oldengott I. M., Boriero D., Schwarz D. J., 2016, JCAP, 08, 054
\bibitem{Oman15} Oman, K.A. et al., 2015, MNRAS, 452, 3650
\bibitem{Padmanabhan} Padmanabhan, N., Finkbeiner, D. P., 2005, Phys. Rev.D, 72, 023508, arXiv:astro-ph/0503486
\bibitem{Picozza} Picozza, P. et al., 2007, Astropart. Phys. 27, 296
\bibitem{Planck} Planck Collaboration, 2015, arxiv:1502.01589
\bibitem{Press} Press W. H., Schechter P., Astrophys. J. 187 (1974) 425–438.
\bibitem{Reimers1997} Reimers D., Kohler S., Wisotzki L., Groote D., Rodriguez-Pascual P., Wamsteker W., 1997, A\&Ap,
 327, 890
\bibitem{Reimers2005} Reimers D., Fechner C., Hagen H.-J., Jakobsen P., Tytler D., Kirkman D., 2005, A\&Ap, 442, 63
\bibitem{Ripamonti07} Ripamonti, E., Mapelli, M., Ferrara, A., 2007, MNRAS, 374, 1067
\bibitem{Salucci} Salucci P., Burkert A., 2000, Astrophys. J. Lett., 537,L9
\bibitem{Seager} Seager S., Sasselov D. D., Scott, D., 1999, ApJ, 523, L1
\bibitem{Schleicher08} Schleicher, D.R.G., Banerjee R., Klassen R., 2008, PRD, 78, 3005
\bibitem{Schleicher09a} Schleicher, D.R.G., Banerjee, R., Klessen, R.S., 2009, PRD, 79, 3515
\bibitem{Schleicher09b} Schleicher, D.R.G., Banerjee, R., Klessen, R.S., 2009, PRD, 79, 043510
\bibitem{Scot09} Scot, P., Fairbairn, M., Edsj\"o, J., 2009, MNRAS, 394, 82
\bibitem{Shull85} Shull, J. M.; van Steenberg, M. E., 1985, Astrophys. J., 298, 268
\bibitem{Shull} Shull M., France K., Danforth C., Smith B., Tumlinson J., 2010, ArXiv e-prints, 1008.2957
\bibitem{Slatyer} Slatyer T. R., Padmanabhan N., Finkbeiner D. P., 2009, Phys. Rev. D, 80, 04
\bibitem{Slatyer13} Slatyer, Tracy R., 2013, Phys. Rev. D, 87, 12
\bibitem{Slatyer16a} Slatyer, Tracy R., 2016, Phys. Rev. D, 93, 2, 023527
\bibitem{Slatyer16b} Slatyer, Tracy R., 2016, Phys. Rev. D, 93, 2, 023521
\bibitem{Smette} Smette A., Heap S. R., Williger G. M., Tripp T. M., Jenkins E. B., Songaila A., 2002, ApJ, 564, 542
\bibitem{Smith12} Smith, R.J., Iocco, F., Glover, S.C.O., Schleicher, D.R.G., Klessen, R.S., Hirano, S., Yoshida, N., 2012, ApJ, 761, 154
\bibitem{Spolyar08} Spolyar, D., Freese, K., Gondolo, P., 2008, PRL, 100, 1101
\bibitem{Steigman} Steigman, G., 2015, PRD, 91, 083538
\bibitem{Valdes10} Valdes, M.; Evoli, C.; Ferrara, A; 2010, 404, 1569
\bibitem{Valdes} Valdés, M.; Evoli, C.; Mesinger, A.; Ferrara, A.; Yoshida, N, 2013, MNRAS, 429,2
\bibitem{Watson} Watson, D. F.; Berlind, A. A.; Zentner, A. R.; 2011, ApJ,738, 22
\bibitem{Weidenspointner} Weidenspointner G., Shrader C.R., Kn\"ödlseder J., Jean P., Lonjou V., Guessoum N., Diehl R., Gillard W., Harris M.J., Skinner G.K., et al., A\&A 450, 1013 
 (2006), arXiv:astro-ph/0601673 
 \bibitem{Yoon08} Yoon, S.-C., Iocco, F., Akiyama, S., 2008, ApJ, 688, 1
\bibitem{ZhengCh} Zheng W., Chiu K., Anderson S. F., Schneider D. P., Hogan C. J., York D. G., Burles S., Brinkmann J., 2004, AJ, 127, 656
\bibitem{ZhengKr} Zheng W., Kriss G. A., Deharveng J.-M., Dixon W. V., Kruk J. W., Shull J. M., Giroux M. L., Morton D. C., Williger G. M., Friedman S. D., Moos H. W., 2004, ApJ, 605, 631
\bibitem{ZhengMe} Zheng W., Meiksin A., Pifko K., Anderson S. F., Hogan C. J., Tittley E., Kriss G. A., Chiu K., Schneider D. P., York D. G., Weinberg D. H., 2008, ApJ, 686, 195

\end{thebibliography}
\end{document}